# Investigating the Optical and Thermodynamic Properties of 2D MoGe$_2$P$_4$ : Potential Material for Photothermal Therapy

Sudipta Saha[a], Arpan Sur[a], Labonno Saha[b] and Md. Kawsar Alam[a]

[a] Department of Electrical and Electronic Engineering, Bangladesh University of Engineering and Technology, Dhaka-1205, Bangladesh

[b] Sir Salimullah Medical College Mitford Hospital, Dhaka-1100, Bangladesh

## Abstract

In this study, we analyzed the optical, thermodynamic and electronic properties of 2D MoGe$_2$P$_4$ from the first principle calculation. 2D MoGe$_2$P$_4$ demonstrates superior optical absorption in the NIR-I biological window (750 nm ~ 1000 nm) with a peak near 808 nm and excellent thermal conductivity (63 Wm$^{-1}$K$^{-1}$). Finite-difference time-domain (FDTD) simulations and Heat simulations demonstrate that 2D MoGe$_2$P$_4$ possesses efficient photothermal conversion under low laser power (0.5 W/cm$^2$) which is operated in 808nm. Theoretical investigation demonstrates, rapid temperature elevation ($\Delta T$ = 24.8 °C) of the 2D MoGe$_2$P$_4$ within two minutes and photothermal stability over multiple laser cycles, achieving temperatures suitable for effective photothermal therapeutic applications. Photothermal therapy (PTT) is an emerging tumor treatment technique that utilizes photothermal agents (PTAs) to convert near-infrared (NIR) light into localized heat for tumor ablation. To enhance biocompatibility, we analyzed the PEGylation of 2D MoGe$_2$P$_4$ nanosheets through molecular dynamics simulation. PEGylation at human body temperature was stable which signifies 2D MoGe$_2$P$_4$'s prospect in therapeutic applications. This research highlights the potential of 2D MoGe$_2$P$_4$ as an emerging material for PTA, establishing a foundation for experimental and clinical trials.

# Introduction

Recently 2D materials has achieved significant attention due to their applications in different research fields such as biosensing, biomedicine, energy storage, nanoelectronics and gas sensing. 2D materials possess enhanced electronic, optical, mechanical and chemical properties depending on their phase, degree of exfoliation, size and stability.[1] These properties of the material significantly improved sensing and biosensing performance of the system in electrochemical and optical mode. Besides these properties makes 2D materials promising candidates for drug delivery, photothermal and photodynamic therapies.[2] Conventional treatment modalities for tumor treatment, such as chemotherapy and radiotherapy, despite their prevalent use in clinical practice, exhibit substantial drawbacks, such as considerable damage to healthy tissues and a high risk of recurrence of tumor cells.[3, 4] These drawbacks highlight the insufficiency of traditional methods in meeting the therapeutic demands for effective tumor treatment fully. Recently, photothermal therapy (PTT) has emerged as a promising alternative to selectively target and destroy tumor cells with reduced side effects.[5, 6] PTT employs a photothermal agent (PTA) that absorbs light, typically in the near-infrared (NIR) range or from other external sources, converting this light energy into heat.[7] This process begins with the excitation of the PTA from its ground state to an excited state,[8] followed by non-radiative decay and relaxation, which releases heat into the surrounding microenvironment.[9] Two-dimensional materials, including graphene, boron nitride, and molybdenum disulfide, have emerged as promising candidates for PTT alongside other inorganic and organic materials.[10] These 2D nanomaterials are suitable for PTT because of their many beneficial qualities, including a high surface area-to-volume ratio[11, 12], biocompatibility,[13] stability in physiological settings, [14] ease of production and functionalization,[15] and high photothermal conversion efficiency.[16]

Graphene and its derivatives have unique physical and chemical properties, making them promising material for PTT-based cancer treatment.[17] However, a common limitation of using graphene for biomedical applications is its poor solubility in water due to its strong $\pi$–$\pi$ bonding.[18] Rasel et al. discovered that boron nitride nanoparticles enter human osteoblast cells in three steps: they first stick to the cell membrane, then start to be absorbed, and finally move inside the cell, showing promise for biomedical uses and PTT.[19] Studies on the toxic effects of boron nitride nanoparticles have shown they are generally biocompatible with various cell types, including pheochromocytoma cells, human neuroblastoma cells, muscle cells, and endothelial cells.[20] Nevertheless, a noted limitation is that these nanoparticles tend to form large, uneven aggregates in aqueous solutions, which could hinder cell uptake.[21] Chou et al. initially documented the potential of utilizing synthesized $MoS_2$ nanosheets as an effective near-infrared absorption agent for in vitro photothermal ablation of cancer cells.[22] Liu et al. enhanced the hydrophilicity of $MoS_2$ nanosheets by modifying them with iron oxide and PEG, creating a platform that combines the photothermal capabilities of TMD nanosheets with biological imaging.[23] Although $MoS_2$ offers an alternative to graphene, it has limitations, including its low absorption of near-infrared (NIR) radiation in biological tissues.[24] Insufficient absorbance may inhibit the considerable temperature elevation necessary to ablate tumor cells during photothermal therapy. MXenes possess excellent absorption in the NIR-II range, which is helpful for in vivo PA imaging and PTT.[25] Nonetheless, nanoparticles operating in the NIR-I are more advantageous for photothermal therapies than those in the NIR-II.[26] In the NIR-I window, water absorption is negligible, making nanoparticles the sole heating source. In contrast, the NIR-II window exhibits water absorption across the entire range, with a coefficient of 0.3 cm$^{-1}$ at 1090 nm, leading to background heating. This reduces the selectivity and efficiency of photothermal therapy, as confirmed by in vivo experiments.[27]

Recent first-principles studies confirm that the two-dimensional $MoSi_2N_4$ family exhibits remarkable piezoelectric properties, high thermal conductivity, exceptional stiffness and promising potential for photocatalytic applications, showcasing its suitability for advanced functional materials.[28] The novel 2D $MoSi_2N_4$ was successfully fabricated experimentally through chemical vapor deposition. Other 2D materials of the $MoA_2Z_4$ family can possibly be synthesized experimentally.[29-31] Among them, $MoGe_2P_4$ has garnered significant attention due to its enhanced optical absorption in the NIR-I window which makes it an ideal material for photothermal therapeutic applications.

In this study, we systematically analyzed the material properties of 2D $MoGe_2P_4$, including its optical, band structure, and thermodynamic characteristics, using density functional theory (DFT). Subsequently, a simulation model was developed using the Finite Difference Time Domain (FDTD) to investigate the interaction of 808 nm laser light, commonly used in medical applications,[26] with 2D $MoGe_2P_4$. The simulation framework assessed light absorption by a photothermal agent, 2D $MoGe_2P_4$, within a tumor environment. $MoGe_2P_4$ demonstrated superior thermal conductivity and reduced temperature rise times compared to other two-dimensional materials. The bioheat transport equation was solved to generate heat maps at different time stamps. Additionally, we examined the effects of pulse laser irradiation on temperature variations, allowing for temperature control and analysis of thermal dynamics. Finally, we assessed the possible toxicity of 2D $MoGe_2P_4$ in a biological context and explored strategies to mitigate any associated risks through our simulation framework.

## Methodology

Ab initio calculations for monolayer $MoGe_2P_4$ are performed using Density Functional Theory (DFT) using the CASTEP[32] module in Material Studio 2020 software. The Generalized Gradient Approximation (GGA) with the Perdew-Burke-Ernzerhof (PBE)[33] exchange-correlation function was employed in this study. All calculations use an ultrasoft pseudo potential with a plane-wave basis set. The cutoff energy was set to 500 eV. The geometry of 2D $MoGe_2P_4$ is relaxed until the energy and force reach the criteria of convergence at $1.0 \times 10^{-6}$ eV and 0.01 eV Å$^{-1}$ between two adjacent steps. A vacuum slab of 25Å was incorporated in the z-direction to avoid the undesired interaction of periodic images, and Grimme's DFTD dispersion correction method was utilized to study the influence of layer-to-layer van der Walls interactions.[34] Structural optimization is performed using a Brillouin zone sampling grid of 13×13×1. To mitigate bandgap underestimation by the PBE algorithm, we used the hybrid functional Heyd–Scuseria–Ernzerhof (HSE06)[35] for band structure calculations. The optical properties were calculated by incorporating spin-orbit coupling (SOC). Phonon is computed using the CASTEP module. An *ab initio* molecular dynamics (AIMD) study was conducted for $MoGe_2P_4$ within the NVT and NVE ensemble. The simulations employed a time step of 1 fs over 2000 steps, using a Nosé–Hoover thermostat[36] to control the temperature at 373 K.

The optical interactions between 2D MoGe$_2$P$_4$ nanoflakes and an incident laser field were analyzed using a three-dimensional FDTD simulation approach. The three-dimensional Maxwell's equations were solved using Ansys Lumerical FDTD software to model the optical response.

The configuration of the 2D MoGe$_2$P$_4$ nanoflakes within a simulated tumor environment was established as shown in **Figure 1(a)**. Besides, **Figure 1(b)** depicts the top view of the simulation environment. The complex refractive indices of the 2D MoGe$_2$P$_4$ nanoflakes were obtained from the first principle study. To represent the surrounding tumor medium at the wavelength of interest, a refractive index of 1.328 was used.[37] The imaginary component of the tumor's refractive index was neglected due to its negligible effect on the simulation outcomes.

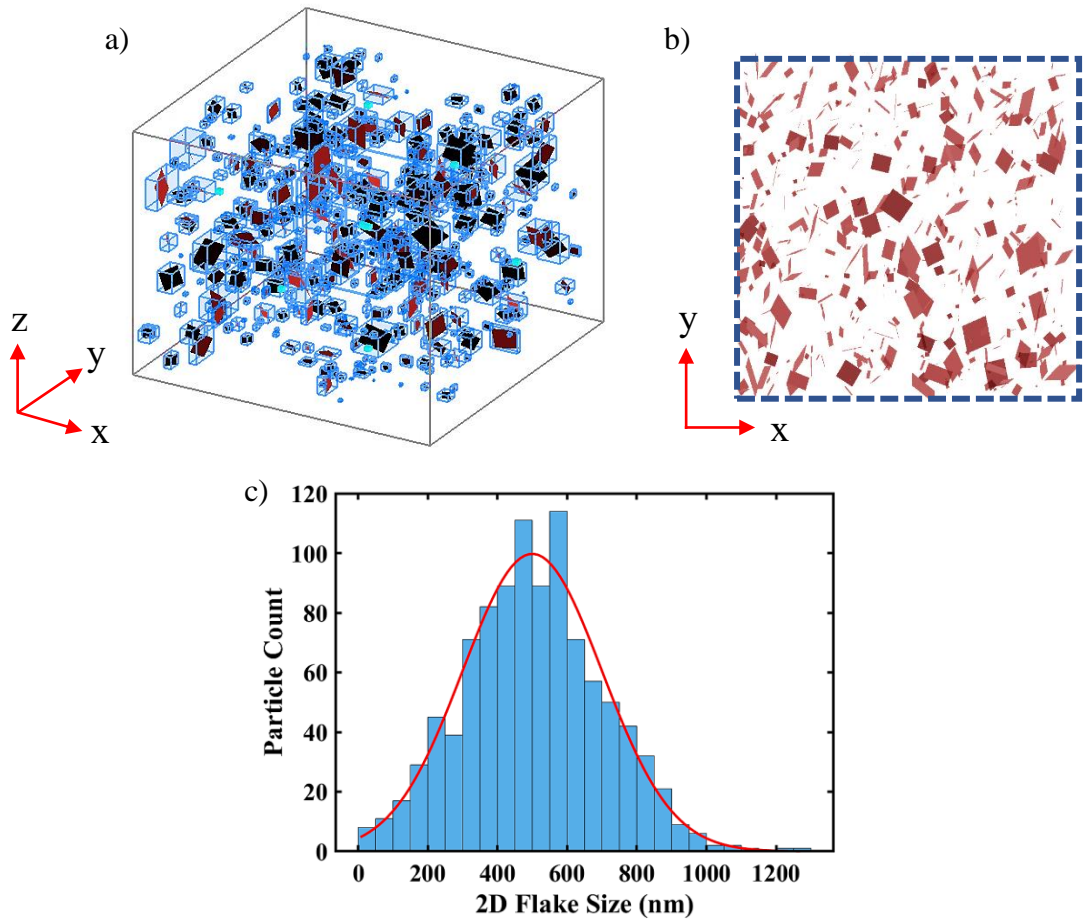

**Figure 1**. 3D FDTD simulation of a bio-cellular environment containing 2D MoGe$_2$P$_4$ nanoflakes. (a) Perspective view showing 2D MoGe$_2$P$_4$ nanoflakes surrounded by a mesh grid. (b) XY-plane cross-section of the distributed 2D MoGe$_2$P$_4$ nanoflakes. (c) Particle count versus 2D MoGe2P4 flake size. The lateral dimensions of each nanoflake are sampled from a Gaussian distribution with a mean of 0.5 µm and a variance of 0.3 µm. The tumor medium was modeled within a cuboid box region measuring $5 \times 5 \times 4$ µm³.

The 2D nanoflakes were randomly positioned and oriented, following a uniform distribution to model their arrangement in a solution. The lateral dimensions of each nanoflake were sampled from a Gaussian distribution with a mean of 0.5 µm and a variance of 0.2 µm as shown in **Figure 1(c)**.[38] The thickness of each nanoflake was fixed at 0.984 nm, based on the lattice thickness in the C direction.

The tumor medium was modeled within a cubic simulation region measuring $2 \times 2 \times 2$ µm³. Periodic boundary conditions were applied in the x- and y-directions, with a perfectly matched layer (PML) boundary condition in the z-direction to prevent reflections. Illumination was provided by a plane wave source at a wavelength of 808 nm, propagating along the z-axis and polarized along the x-axis. Although the experimental setup employed a Gaussian laser beam, the smaller size of the simulated region relative to the experimental interrogation area allowed for a plane wave approximation.[39]

A three-dimensional frequency-domain field and power monitor were used to measure the absorbed optical power across varying input laser intensities. This allowed for a comprehensive analysis of the interaction between the laser and the 2D nanoflakes.

The thermal response of the 2D MoGe$_2$P$_4$ nanoflakes, following their interaction with a laser field, was determined by solving Pennes' bioheat equation: [40]

$$\rho c_p \frac{\partial T}{\partial t} = \nabla \cdot (k \nabla T) + Q_b + Q_m + Q_s \quad (1)$$

Where ρ represents tissue density (kg/m³), $c_p$ is the tissue's specific heat (J/kg·K), and $(k)$ Is the thermal conductivity (W/m·K). The term $Q_b$ Accounts for the heat exchanged due to blood perfusion, formulated as: [41]

$$Q_b = \omega_b \rho_b c_b (T_b - T) \quad (2)$$

Where, $\omega_b$ denote blood perfusion rate, blood density, blood-specific heat, and blood temperature. Additional terms include $Q_m$, the volumetric metabolic heat generation rate, and $Q_s$, The rate of absorbed laser power per unit volume of tissue.[42]

The bioheat equation was modeled and solved using the Ansys Lumerical Heat Transport (HEAT) solver, which provided the temperature distribution across the simulation domain. The thermal properties of the tumor cells and the 2D MoGe$_2$P$_4$ nanoflakes used in the simulations are presented in **Table 1**.

Table 1: Thermal Properties of Tumor Cells and 2D MoGe$_2$P$_4$

| Data | Tissue [41] | MoGe$_2$P$_4$[28] |
|---|---|---|
| **Density** | 1100 kg/m³ | 5890 kg/m³ |
| **Specific heat** | 4200 J/kg·K | 371.58 J/kg·K |
| **Thermal Conductivity** | 0.55 W/(m·K) | 63 W/(m.K) |

The thermal simulations employed the same simulation region and nanoflake configuration as in the FDTD optical simulation. Blood perfusion effects were modeled by applying a constant temperature boundary condition of 37°C to the $x_{\min}, x_{\max}, y_{\min}, y_{\max}$, and $z_{\min}$ surfaces of the simulation domain. Metabolic heat generation was modeled with a uniform heat source. For the 2 × 2 × 2 μm³ simulation region, the input power was calculated to be 8.728×10⁻¹⁵ W based on a background body metabolic heat of 1091 W/m³ .[41]

The spatial heat generation profile from absorbed light $Q_s$ As imported from the FDTD simulation and applied as a heat source. The convective heat transfer between the tumor cells and 2D nanoflakes was modeled with a heat transfer coefficient. $h = 5$ W/(m²·K), while maintaining an ambient temperature of 37°C .[43] A temperature monitor was implemented to record the resultant temperature distribution across the simulation domain.

Singh et al. highlighted that the effectiveness of hyperthermia treatment depends on sufficient exposure time to achieve therapeutic effects.[44] The extent of irreversible thermal damage to biological tissues during hyperthermia is commonly evaluated by integrating the Arrhenius equation over time, as described by Welch[45]

$$\Omega(t) = \int_0^t A \exp\left(-\frac{E_a}{RT(t')}\right) dt' \tag{3}$$

Where, $\Omega(t)$ is the damage parameter, $E_a$ is the activation energy, $A$ is the frequency factor, $R$ represents the gas constant 8.314 J/(mol·K), $T$ denotes the absolute tissue temperature, and $\Omega$ indicates the extent of thermal damage. When $\Omega$ exceeds 1, the tissue is considered permanently damaged.[46] In this study, the Arrhenius parameters of skin, activation energy $E_a = 7.82 \times 10^5$ J/mol and frequency factor $A = 2.185 \times 10^{124}$ $s^{-1}$ were adopted to evaluate thermal damage.[47]

## Result and Discussion

### A. Benchmarking the Simulation Framework

To establish a reliable baseline for the simulations, the geometry and lattice parameters of pristine MoGe₂P₄ were first computed. This initial step provides a benchmark for the simulation environment. The geometry-optimized monolayer MoGe₂P₄ exhibits a hexagonal lattice structure, as depicted in **Figures 2(a) and 2(b)**, with lattice parameters a = b = 3.54 Å. The bond distances within the structure include a Mo–P bond length of 2.48 Å and a P–Ge bond length of 2.53 Å. The thickness of the MoGe₂P₄ monolayer along the C-axis is measured to be 9.48 Å, as illustrated in **Figure 2(c).** These structural parameters are consistent with those reported in the literature.[28, 48, 49]

For accuracy, the electronic band structure of 2D MoGe₂P₄ is calculated using HSE-06 functional methods. As depicted in **Figure 2(d),** 2D MoGe₂P₄ is a direct bandgap material with a band gap of 0.89 eV. The valence band maxima (VBM) and conduction band minima (CBM) are at the K point. The electronic properties also match previously reported values.[28, 48, 49]

To ensure the stability of 2D MoGe₂P₄, phonon calculation was done as presented in **Supplementary Section A**. No negative frequencies were observed in **Figure S1,** which validates that the structure is stable. Besides, the literature suggests that 2D MoGe₂P₄ satisfies Born stability conditions ensuring the mechanical stability of the structure.[48] So, the monolayer is dynamically and mechanically stable.

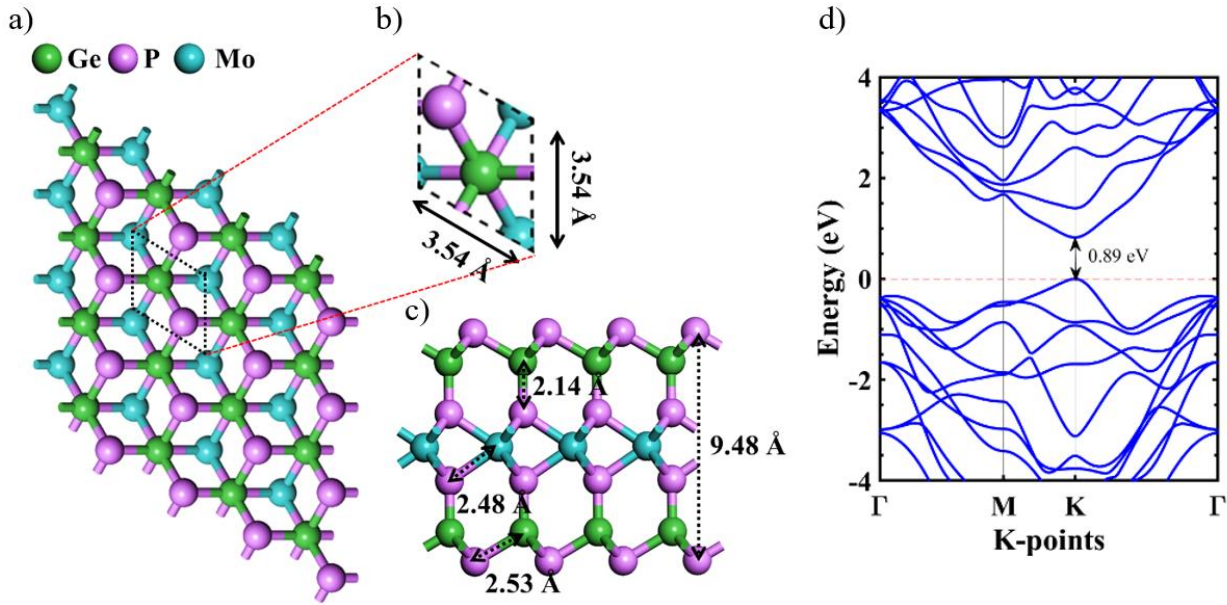

**Figure 2.** The hexagonal lattice structure of 2D MoGe$_2$P$_4$: (a) a top view, (b) a close-up of the unit cell with lattice parameters a = b = 3.54Å, (c) a side view, (d) the electronic band structure plotted along the Γ-M-K-Γ path. In the lattice structure, the atoms are color-coded: green for Ge, purple for P, and sky blue for Mo.

## B. Optical Properties

The absorption coefficient can be derived from the complex dielectric constant, $\epsilon = \epsilon_1 + i\epsilon_2$. The imaginary part ($\epsilon_2$) of the dielectric constant ($\epsilon$) can be calculated using the following formula using the electrical band structure,[50]

$$\epsilon_2(q \to O_{\hat{u}}, \hbar\omega) = \frac{2e^2\pi}{\Omega\epsilon_0} \sum_{k,v,c} |\langle \Psi_k^c | \boldsymbol{u} \cdot \boldsymbol{r} | \Psi_k^v \rangle|^2 \; \delta(E_k^c - E_k^v - \hbar\omega) \qquad (4)$$

Where, **u** is the vector defining polarization of incident light, $\langle \Psi_k^c | \boldsymbol{u} \cdot \boldsymbol{r} | \Psi_k^v \rangle$ is the matrix element. The conduction band energy is $E_k^c$, the valence band energy is $E_k^v$ at the wave number is k, respectively. The angular frequency of the electron is ω, the charge is e, the momentum operator is **u.r**, and ℏ is the reduced Planck's constant. The real part of the dielectric constant can be found using the Kramers – Kronig transformation [51]

$$\epsilon_1(\omega) = 1 + \frac{2}{\pi} P \int_0^\infty \frac{\epsilon_2(\omega^*)\omega^*}{\omega^{*2} - \omega^2} d\omega^* \qquad (5)$$

Where, P is the principal value of the integral. The optical absorption coefficient for 2D MoGe$_2$P$_4$ is calculated as a function of dielectric constants as follows:[52]

Page **7** of 23

$$\alpha(\omega) = \frac{4\pi\kappa(\omega)}{\lambda\sqrt{2}} \left( \sqrt{\epsilon_1^2(\omega) + \epsilon_2^2(\omega)} - \epsilon_1(\omega) \right)^{\frac{1}{2}} \quad (6)$$

Where $\epsilon_1$ and $\epsilon_2$ are the real and imaginary parts of the dielectric function, κ is the extinction coefficient, and λ is the wavelength.

Refractive index (n) and extinction coefficient (k) can be obtained from the dielectric constant by following the formula:[52]

$$n(\omega) = \left[ \sqrt{\epsilon_1^2(\omega) + \epsilon_2^2(\omega)} + \epsilon_1(\omega) \right]^{\frac{1}{2}} / \sqrt{2} \quad (7)$$

$$k(\omega) = \left[ \sqrt{\epsilon_1^2(\omega) + \epsilon_2^2(\omega)} - \epsilon_1(\omega) \right]^{\frac{1}{2}} / \sqrt{2} \quad (8)$$

The absorption coefficient of 2D MoGe$_2$P$_4$ and typical human tissue[26] is shown in **Figure 3(a).** Notably, the absorption spectrum shows a peak for 2D MoGe$_2$P$_4$ around 812.55 nm which is close within the NIR-1 window, with an additional absorption peak also appearing in the NIR-2 window. In typical human tissue, optical absorption is primarily influenced by the absorption bands of key components (such as water and hemoglobin) and the inherent scattering effects of the natural structure of the tissue. Light absorption strongly vanishes for tissue in this NIR-I and NIR-II spectral region. The use of specific laser sources working in the biological windows for photothermal therapy not only reduces the non-selective heating of healthy tissue but also allows for deep-tissue treatments. Indeed, this is due to the extended optical penetration depth obtained from the simultaneous minimization of scattering and absorption processes.[53] Operating within these two biological windows enables the optical excitation of light-activated heating nanoparticles positioned within non-superficial tumors.[54, 55] This approach holds the potential for developing photothermal treatments that target deep tissue tumors through light-induced heating of nanoparticles. Realistically, however, the maximum optical penetration depth achievable in human tissue is limited to only a few centimeters.[53] Photothermal treatment of deeper tumors can still be achieved by combining optical fiber delivery with endoscopic techniques.[10]

Refractive index (n) and extinction coefficient (k) with respect to wavelength are provided in **Figure 3(b). Figure 3(c)** presents a comparative analysis of the absorption range of MoGe$_2$P$_4$ with different heating nanoparticles employed in photothermal therapy (PTT), including gold nanospheres (GNSph), gold nanorods (GNRs), neodymium-doped nanoparticles (Nd:NPs), carbon nanotubes (CNTs), organic nanoparticles (O:NPs), and gold nano stars (GnSs).[26] It can be observed that most of the heating nanoparticles operate within the first biological window (NIR-I), while only a few, specifically gold nanorods (GNRs) and carbon nanotubes (CNTs), are effective in the second biological window (NIR-II). Notably, 2D MoGe$_2$P$_4$ exhibits a broad absorption spectrum spanning the visible range, along with substantial absorption within both the NIR-1 and NIR-2 windows. This extensive absorption profile highlights MoGe$_2$P$_4$ as a promising candidate for photothermal therapy applications.

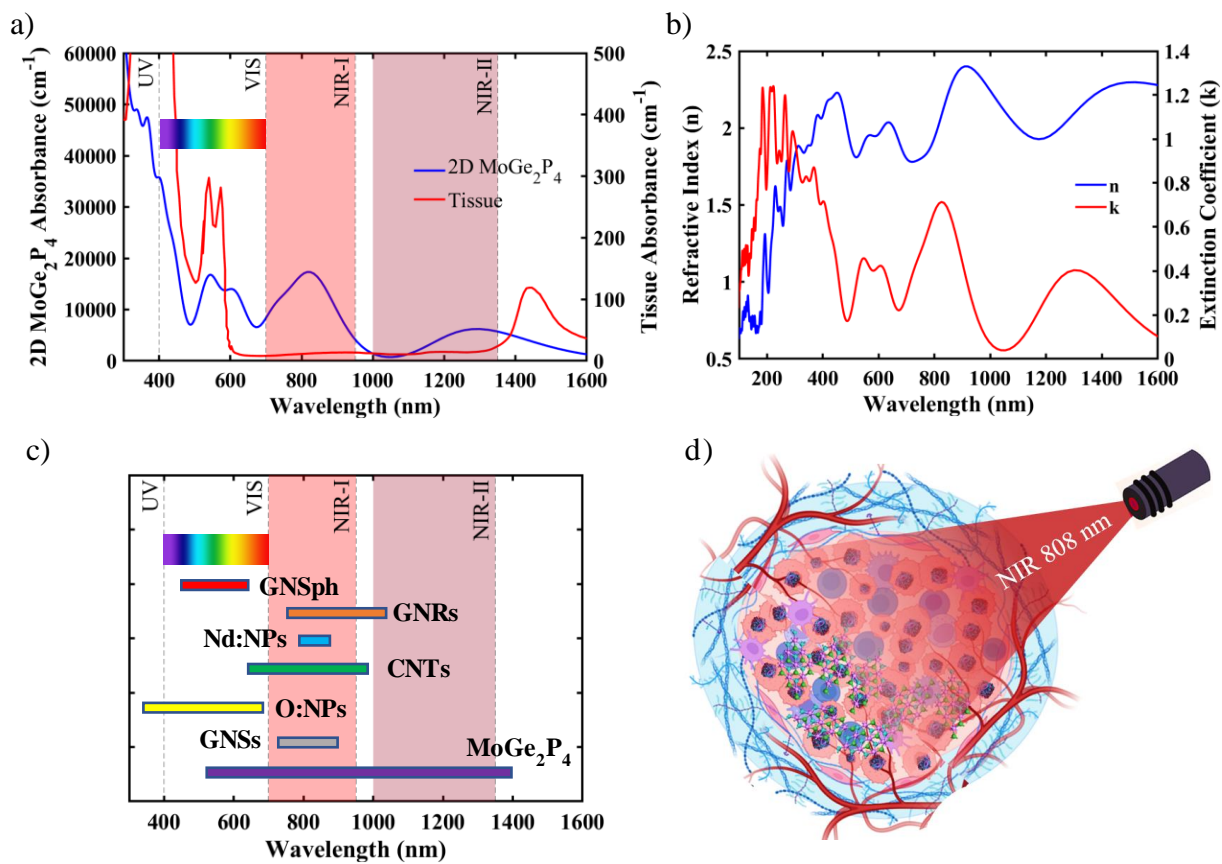

**Figure 3.** Optical characteristics of 2D $MoGe_2P_4$: (a) absorption spectrum $MoGe_2P_4$ vs typical human tissue, (b) refractive index and extinction coefficient, (c) comparison of absorption range with various heating nanoparticles gold nanosphere (GNSph), gold nanorods (GNRs), neodymium-dopes nanoparticles (Nd:NPs), carbon nanotubes (CNTs), organic nanoparticles (O:NPs), gold nano star (GnSs) with our work ($MoGe_2P_4$) for PTT, and (d) mechanism of PTT utilizing an 808 nm Laser. The extended absorption range of $MoGe_2P_4$ highlights its superior potential for PTT, as it covers broader NIR wavelengths and may offer better efficiency and versatility in targeting deep tissues.

**Figure 3(d)** depicts a mechanism in PTT for treatment, illustrating near-infrared (NIR) light at a wavelength of 808 nm to target a tumor. At the center, the tumor is represented by clusters of dark circular shapes symbolizing tumor cells, surrounded by a network of red, branch-like structures resembling blood vessels supplying the tumor with nutrients. Positioned on the right, a laser device emits a red beam labeled "NIR 808 nm" directly onto the tumor area. Within the tumor, small patterned structures appear, representing photothermal agents, which are 2D $MoGe_2P_4$ nanoflakes designed to absorb NIR light and convert it into heat. This localized heating effect, when activated by the NIR laser, has the potential to damage cancer cells selectively. For photothermal therapy applications, utilizing heating nanoparticles in the NIR-1 is generally more advantageous than those operating in the NIR-2 window. This preference arises from the fact that the NIR-1 window is characterized by negligible absorption of water in this window, which means that the primary source of heating would be the nanoparticles themselves.[56]

## C. Thermodynamic Properties

The thermal stability of the 2D MoGe$_2$P$_4$ monolayer was assessed through ab initio molecular dynamics simulations conducted in NVT ensembles. In the NVT ensemble, the number of atoms (N), volume (V), and temperature (T) were held constant with a Nosé–Hoover thermostat maintaining the temperature at 373 K, using a time step of 1 fs for 2000 steps. Results from these simulations indicate that the energy fluctuations, which are benchmark for the thermal stability of the monolayer, remained minimal within 0.0278 eV, respectively, during the MD computations in the NVT. Notably, the MoGe$_2$P$_4$ monolayer maintained its structural stability throughout the 2000-step simulation (with a 1 fs time step) at 373 K.

With the calculation of phonon properties, the thermodynamic enthalpy H, the free energy F, and the entropy S at finite temperature can be computed. The vibration contribution to the free energy is derived as follows:

$$H(T) = E_{\text{tot}} + \frac{1}{2} \int g(\omega) \hbar \omega \, d\omega + \int \frac{\hbar \omega}{e^{\frac{\hbar \omega}{k_B T}} - 1} g(\omega) \, d\omega \tag{9}$$

$$F(T) = E_{\text{tot}} + \frac{1}{2} \int g(\omega) \hbar \omega \, d\omega + k_B T \int g(\omega) \ln\left(1 - e^{\frac{-\hbar \omega}{k_B T}}\right) d\omega \tag{10}$$

$$S(T) = k_B \left[ \int \frac{\hbar \omega}{k_B T} \frac{g(\omega) d\omega}{e^{\frac{\hbar \omega}{k_B T}} - 1} - \int g(\omega) \ln\left(1 - e^{\frac{-\hbar \omega}{k_B T}}\right) d\omega \right] \tag{11}$$

Where, $k_B$ is the Boltzmann constant, $\hbar$ is the reduced Planck's constant, $E_{\text{tot}}$ is total electronic energy at 0K, and $g(\omega)$ is the phonon density of states.

The lattice contribution to heat capacity can be obtained from the following equation: -

$$C_v(t) = k_B \int \frac{\left(\frac{\hbar \omega}{k_B T}\right)^2 \exp\left(\frac{\hbar \omega}{k_B T}\right)}{\left(\exp\left(\frac{\hbar \omega}{k_B T}\right) - 1\right)^2} F'(\omega) \, d \tag{12}$$

Debye temperature can be obtained [57]

$$C_v^D(T) = 9 N k_B \left(\frac{T}{\Theta_D}\right)^3 \int_0^{\frac{\Theta_D}{T}} \frac{x^4 e^x}{(e^x - 1)^2} dx \tag{13}$$

Where N is the number of atoms per cell and $\Theta_D$ is the Debye temperature.

Thermal conductivity for MoGe$_2$P$_4$ is found to be 63 W/(m.k), which is isotropic.[28] Heat capacity and thermal conductivity parameters are used for heat dissipation and thermal analysis.

A classical MD simulation was conducted at up to 10ps at body temperature T=27°C to visualize the stability, as provided in **Movie S1.** We further incorporated water molecules in the MD simulation to resemble the natural bio environment, as shown in **Movie S2.** In each case, 2D MoGe$_2$P$_4$ was stable.

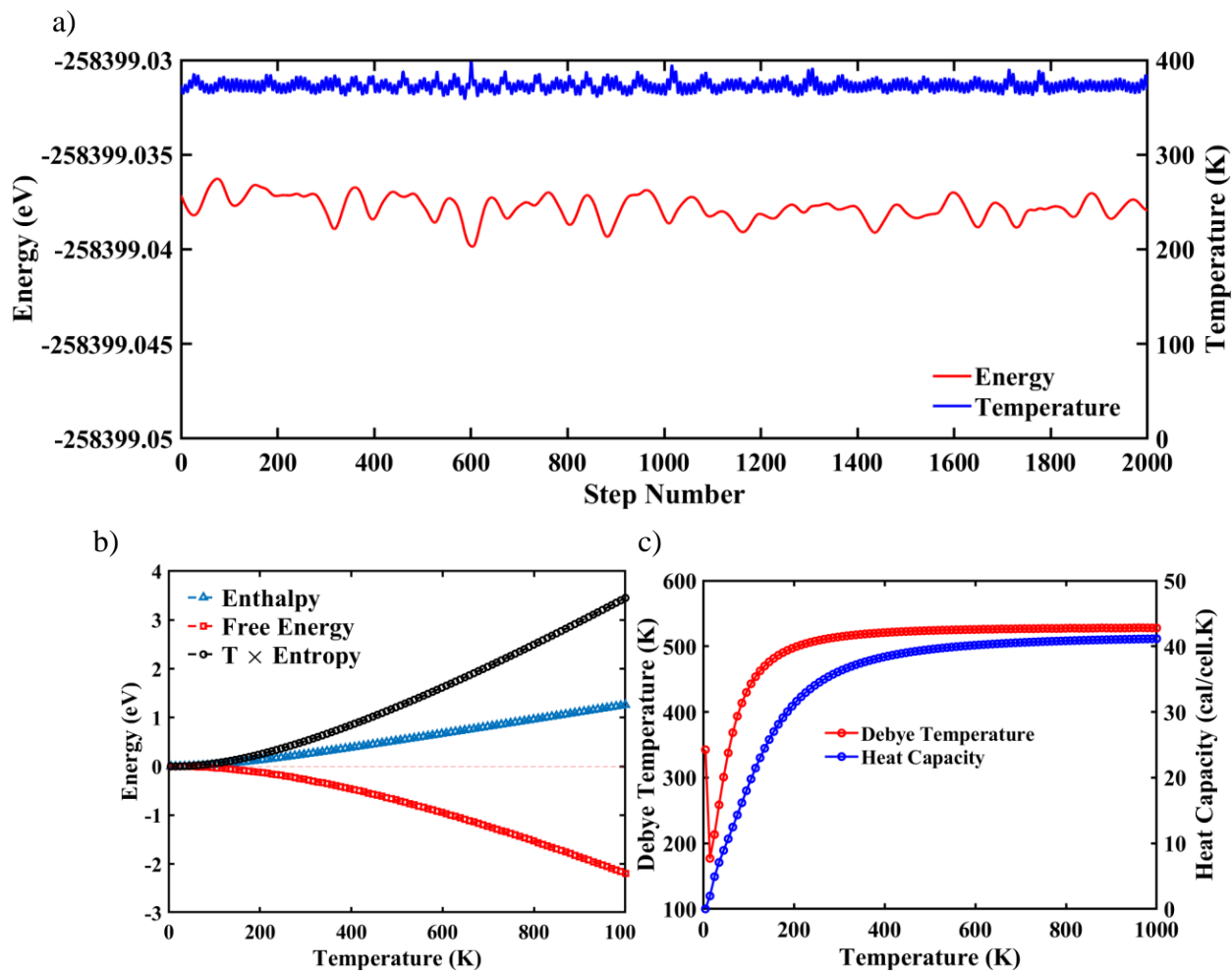

**Figure 4.** (a) The fluctuation of total energy and temperature in NVT ensemble from 0 to 2 ps (time step of 1 fs in 2000 steps), during ab initio molecular dynamic simulation at the temperature of 373 K, (b) Thermodynamic properties for 2D MoGe$_2$P$_4$, (b) Temperature dependence of Debye temperature and Heat capacity.

D. **Photothermal Response of 2D MoGe$_2$P$_4$**

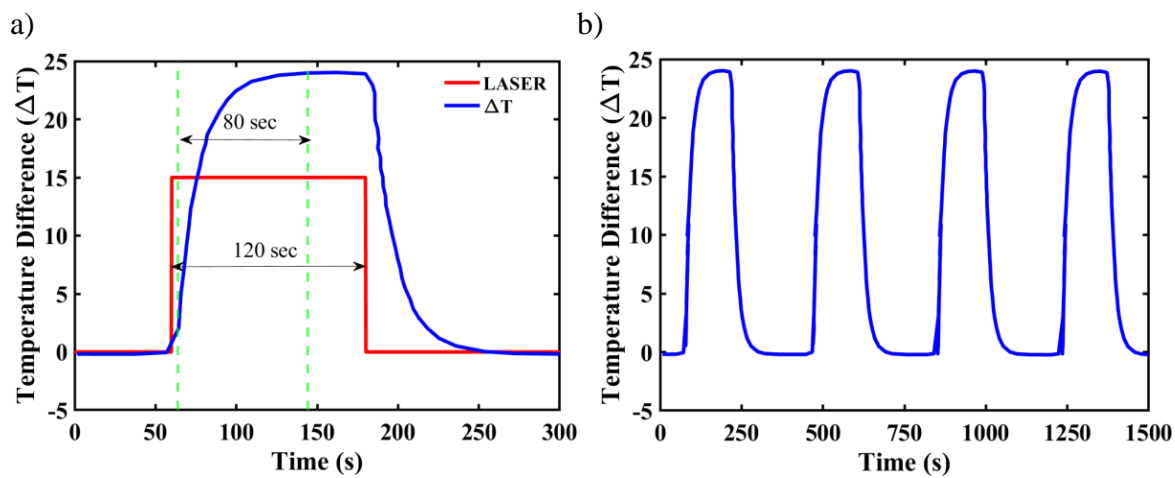

**Figure 5.** Photothermal Response of 2D MoGe$_2$P$_4$ at 808nm Laser irradiation of 0.5 Wcm$^{-2}$ (a) temperature difference observed in tumor with respect to body temperature (37 °C) versus time, demonstrating a temperature rise over an 80-second interval under 808 nm laser exposure with pulse duration of 120 seconds. (b) Extended temperature difference versus time over four on/off laser cycles, highlighting the material's thermal response and photothermal stability across multiple irradiation cycles. The maximum temperature difference is observed to be 24.8°C which is achieved after two minutes of laser irradiation.

**Figure 5** illustrates the photothermal response of 2D MoGe$_2$P$_4$ nanoflakes in a bio-cellular environment. These nanoflakes are configured as shown in **Figure 1(a)**. A laser excitation of 0.5 W/cm² at an 808 nm wavelength was applied for 120 seconds. The ambient temperature of the simulation ground was maintained at 310 K, equivalent to standard human body temperature. **Figure 5(a)** shows the temperature change resulting from laser excitation. The red curve represents the laser pulse amplitude, while the blue curve depicts the maximum temperature change in the system. The temperature rise time, denoted by green dotted lines, is measured to be 80 seconds. Beyond this point, a saturation phase is observed in the temporal temperature profile, indicating the system has reached its maximum temperature.

**Figure 5(b)** presents the temporal temperature profile across four on/off laser cycles. Each cycle exhibits consistent temperature rise and fall patterns, demonstrating a stable thermal response under repeated irradiation. For the 120-second irradiation at 0.5 W/cm², the system achieves a final temperature of 61.8°C, corresponding to a temperature increase of ΔT = 24.8°C.

For temporal analysis of the temperature profile, x-y, y-z, and z-x planes were selected, as shown in **Figure 6.**

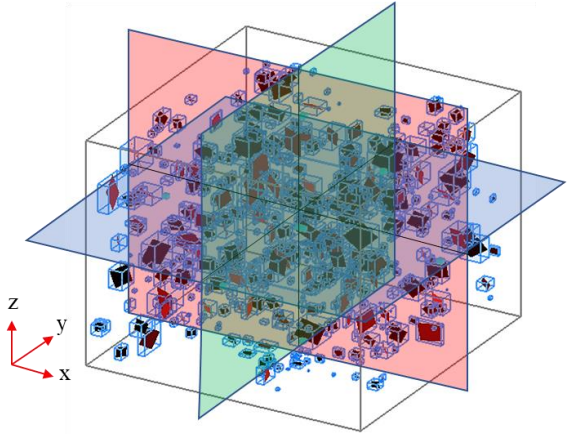

**Figure 6.** 2D cross sections of the simulation ground for temperature analysis. Blue, green, and red planes refer to x-y, y-z, and z-x planes, respectively.

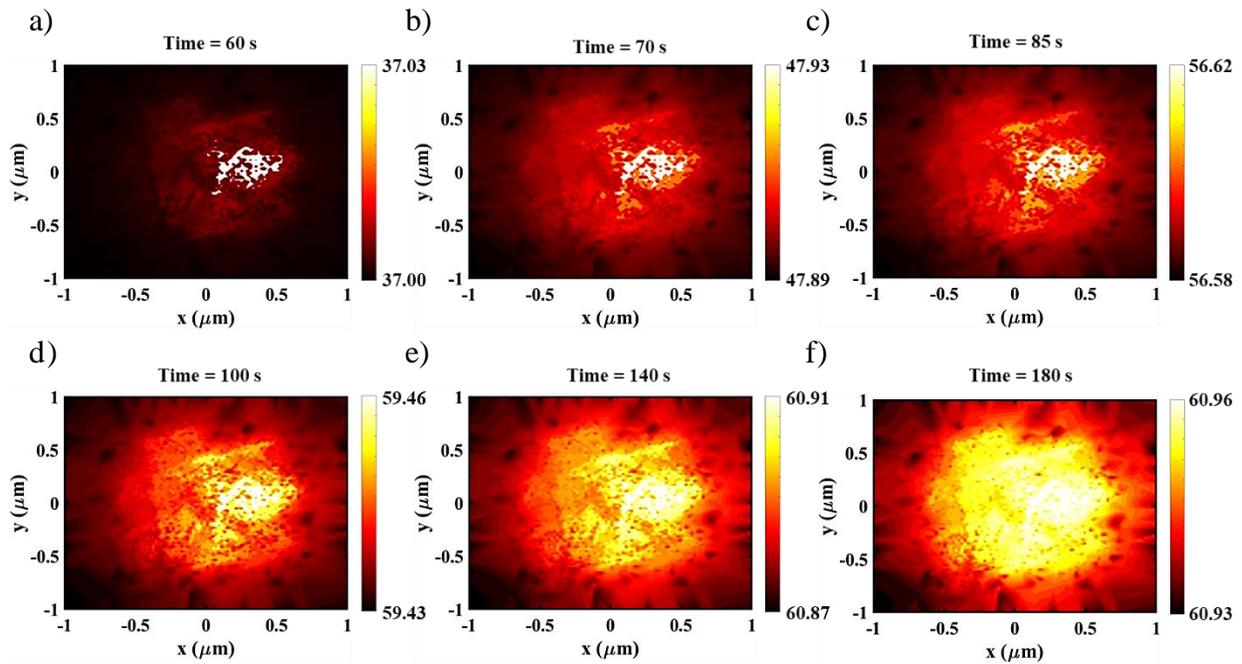

**Figure 7.** 2D Spatial Temperature Distribution in the x-y Plane Over Time: Profiles at (a) 60 seconds, (b) 70 seconds, (c) 85 seconds, (d) 100 seconds, (e) 140 seconds, and (f) 180 seconds. Temperature rises ~30°C at first 30 sec of laser irradiation. Then rise ~1.5°C in the next 80 seconds. Results indicate a rapid initial heating phase, followed by a slower, stabilized temperature increase

**Figure 7** presents a 2D spatial x-y temperature profile for various time intervals. From **Figure 5(a)**, it is evident that the temperature increases sharply immediately after the application of the laser pulse. Multiple temperature profiles were captured in **Figure 7** at shorter intervals during the initial phase to verify this behavior. The results show that the temperature rises by nearly 30 °C within the first 40 seconds. Subsequently, the temperature increases by only 1.5 °C over the next 80 seconds. This rapid initial temperature rise is attributed to the strong optical absorption and efficient thermal coupling of the 2D nanoflakes with the system. The high-temperature region observed at the middle-right position in the x-y plane can be linked to the relatively higher nanoflake density in that area, as indicated in **Figure 7**. Temperature profile analysis in the other two planes is described in **Supplementary Section B.**

**Figures S2 and S3** illustrate the 2D spatial x-z and y-z temperature profiles, respectively. Both figures reveal a high-temperature region at the upper z positions. This phenomenon is due to the downward propagation of the light field. Nanoflakes located at the upper z positions interact more intensely with the light compared to those at lower z positions, resulting in higher optical power absorption and elevated temperatures in the upper region.

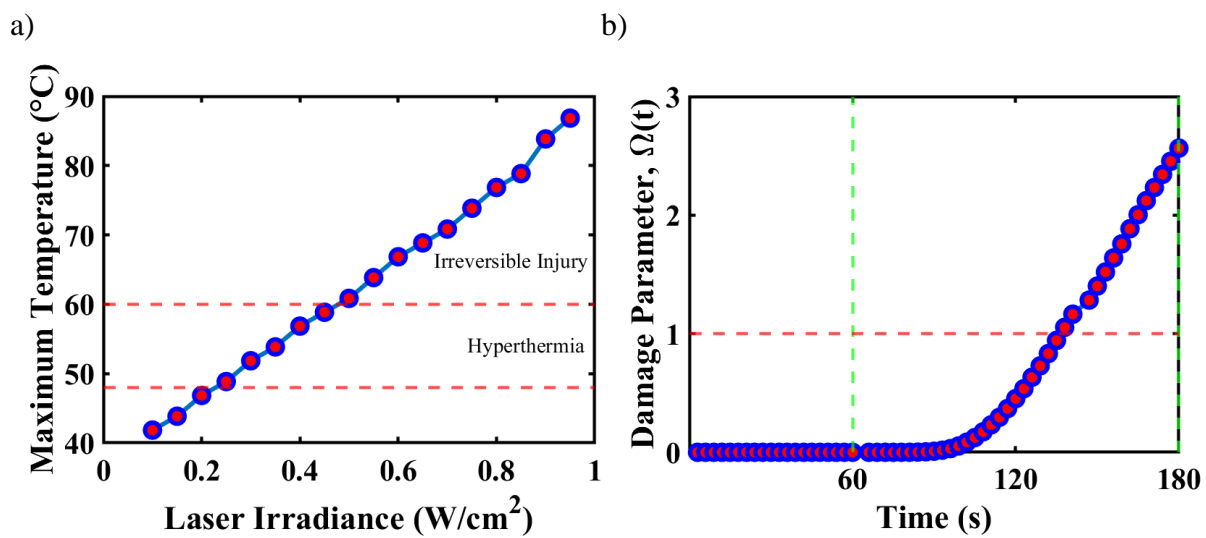

**Figure 8.** (a) Maximum temperature achieved as a function of laser power. The temperature range of 41°C to 48°C corresponds to the hyperthermia zone, while 48°C to 60°C represents the range for irreversible tumor cell injury. (b) Thermal damage distributions are calculated using the Arrhenius equation over time. The damage parameter $\Omega(t)$ exceeds one at 135 seconds, indicating significant thermal injury.

In **Figure 8 (a)**, the maximum achieved temperature after 120 seconds of excitation is plotted with respect to the laser irradiance. A linear relation is found between achieved temperature and laser power. Fitting the data using a straight line gives us a 51.46 K/ (W/cm²) slope. This demonstrates a consistent temperature tunability in response to varying laser power. At a laser intensity of 0.25–0.5 W/cm², the system is well-suited for hyperthermia-based tumor treatment, though prolonged exposure (>60 minutes) is necessary to induce irreversible tumor cell damage. In contrast, when the laser power exceeds 0.5 W/cm², irreversible tumor cell damage occurs more rapidly. From **Figure 8(b)**, it can be observed that the damage parameter exceeds 1 at 135 sec.

**Table 2** provides a comparative analysis of various 2D materials based on their laser power, temperature rise time, and maximum temperature $T_{max}$ achieved under irradiation. In this study, the 2D material under investigation achieved a $T_{max}$ of 61.8°C with a laser power of 0.5 W/cm² at an 808 nm wavelength and a rapid temperature rise time of 80 s. This performance surpasses other materials, such as $MoS_2$ (57.9°C at 1 W/cm$^2$) and SnTe (57.0°C at 1.5 W/cm$^2$) under similar conditions. The material demonstrates superior heating efficiency and faster response compared to BP, $MoS_2$/$Bi_2S_3$, and $Mo_2C$. Materials irradiated with 1064 nm lasers, such as $Nb_2C$ and silicene, exhibit lower $T_{max}$ despite higher laser powers and longer rise times, emphasizing the enhanced photothermal properties of the material in this work.

**E. Application Description**

The Intratumoral (IT) injection is a targeted approach primarily utilized for superficial skin cancers such as basal cell carcinoma, squamous cell carcinoma, and melanoma, as well as accessible solid tumors like breast cancer and superficial lymphomas.[58-60] It can also be employed for localized prostate cancer when the tumor is directly accessible.[61] This method is designed to maximize local concentrations of photothermal agents at the tumor site, thereby reducing systemic exposure and enhancing localized therapeutic efficacy.

2D $MoGe_2P_4$ shows promise as a photothermal agent (PTA). When exposed to an NIR laser, typically at 808 nm with 0.5 Wcm$^{-2}$, $MoGe_2P_4$ absorbs the laser energy and converts it into localized heat. This heat raises the temperature of tumor cells to 61.8°C, inducing irreversible damage, which disrupts cellular functions and leads to apoptosis or necrosis. To effectively and selectively eliminate tumor cells, inducing maximum apoptosis is crucial.[62] This ensures that healthy tissues remain unharmed. In our study, we can regulate laser power and irradiation time to precisely control apoptosis induction as discussed in the previous section. **Figure 9** shows the proposed application process of 2D $MoGe_2P_4$ for tumor cell ablation.

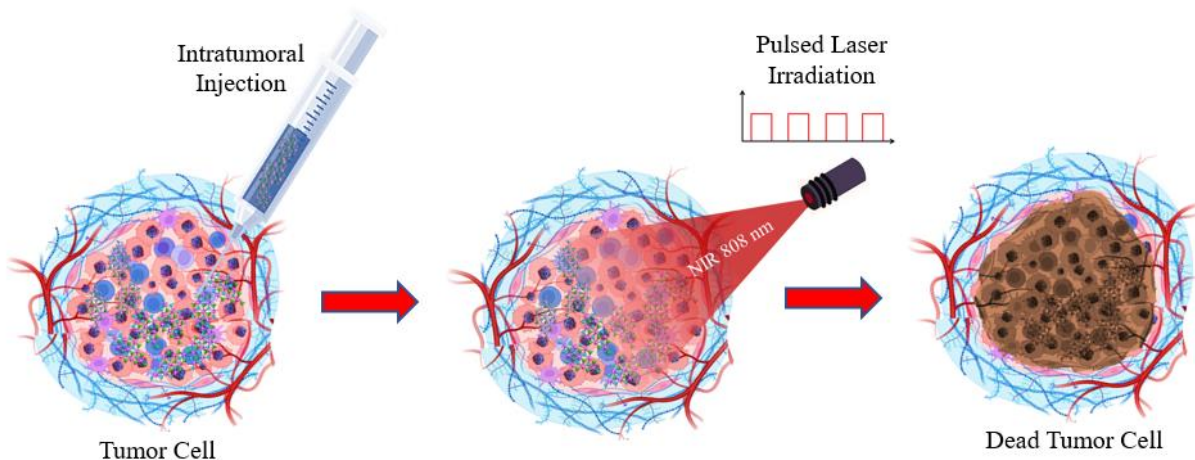

**Figure 9.** Application of 2D $MoGe_2P_4$ via intratumoral injection followed by 808 nm pulsed laser irradiation for complete tumor cell ablation.

## F. Toxicity Analysis

Light-activated nanoparticles must exhibit high absorption within biological optical windows and efficient light-to-heat conversion for effective photothermal therapies, enabling treatment with low-power lasers. Selectivity and minimal side effects are crucial, requiring non-toxic properties without light. Easy functionalization allows precise tumor targeting[63] and good solubility in biocompatible liquids supports prolonged circulation for effective tumor access.[64] Together, these properties make nanoparticles suitable for safe, targeted, and efficient photothermal treatments.

However, long-term safety, specifically concerning the biodegradation and excretion of 2D nanomaterials, remains critical. Many 2D nanomaterials, due to high crystallinity and few structural defects, resist degradation in physiological conditions. Nonetheless, certain types can be rationally designed for biodegradability—2D $MoGe_2P_4$ comprising molybdenum (Mo), germanium (Ge), and phosphorus (P). To find out about the toxicity of Mo, Hao et al. demonstrated that $MoS_2$ degrades into water-soluble Mo (VI) oxide species (e.g., $MoO_4^{2-}$), which are efficiently excreted through renal and fecal pathways, suggesting a low retention and toxicity profile.[65] This efficient clearance suggests a low level of retention and toxicity for Mo. As seen in BP (black phosphorus) nanodots, Phosphorus can degrade into non-toxic phosphate and phosphonate in water and oxygen, facilitating safe clearance without notable toxicity.[66] Xianwei et al. showed that Ge-containing nanostructures (e.g., Ge/GeP nanosheets) exhibit high cellular uptake and biocompatibility with minimal cytotoxicity. Furthermore, animal models' histological and blood biochemical analyses show no significant renal, hepatic, or overall tissue toxicity.[67] These findings indicate that Ge in $MoGe_2P_4$ could contribute to biocompatibility, with efficient clearance from the body and minimal toxicity in normal tissues.

Functionalization strategies can be employed to enhance biocompatibility and optimize the circulation time of 2D $MoGe_2P_4$ nanosheets in biomedical applications, as they have been proven effective for other 2D nanomaterials. One prominent method, PEGylation, involves coating $MoGe_2P_4$ with polyethylene glycol (PEG). This non-toxic, water-soluble polymer has been shown to significantly improve the biocompatibility and dispersibility of nanosheets in physiological environments in previous studies.[68, 69] As shown in **Figure 10 and Supplementary Section C,** PEGylated 2D $MoGe_2P_4$ nanosheets are likely to remain stable. This stability is crucial, especially given 2D $MoGe_2P_4$'s promising role in therapeutic applications where long-term circulation and biocompatibility are required.

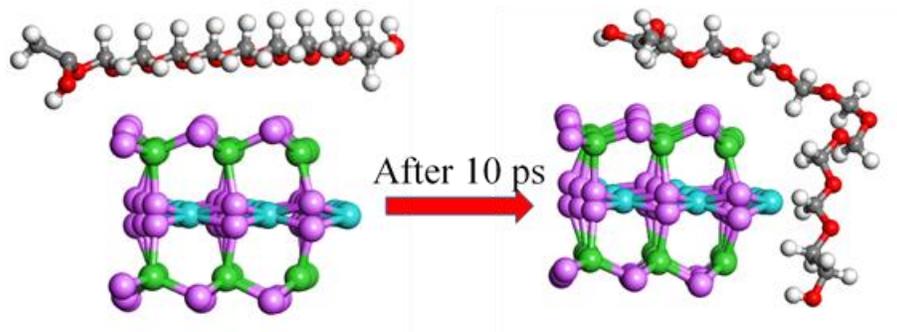

**Figure 10.** Interaction of PEG and 2D MoGe$_2$P$_4$ after 10 ps at body temperature, T = 27°C.

**Table 2:** Comparison of existing 2D Materials based on laser power, thermal conductivity, temperature rise time, and maximum temperature achieved

| 2D material | Laser Wavelength (nm) | Laser power (W/cm$^2$) | Thermal Conductivity (W/m.k) | Temperature Rise Time (s) | T$_{max}$ (°C) |
|---|---|---|---|---|---|
| **This work** | 808 | 0.5 | 63 | 80 | 61.8 |
| **MoS$_2$** [70] | 808 | 1 | 38.73[71] | 250 | 57.9 |
| **SnTe** [72] | 808 | 1.5 | 12.40[73] | 300 | 57.0 |
| **BP** [74] | 808 | 0.3 | 51.3[75] | 300 | 55.2 |
| **MoS$_2$/Bi$_2$S$_3$** [76] | 808 | 0.8 | - | 300 | 57.2 |
| **Mo$_2$C** [77] | 808 | 1 | 5.13[78] | 200 | 52.5 |
| **Nb$_2$C** [79] | 1064 | 1.5 | 6.00[80] | 350 | 52.3 |
| **FePS$_3$** [81] | 1064 | 1 | - | 300 | 53.0 |
| **Silicene** [82] | 1064 | 1.5 | 9.40[83] | 1000 | 52.3 |

In this study, we demonstrated the exceptional photothermal conversion efficiency of the 2D material MoGe$_2$P$_4$, which exhibited a remarkably low-temperature rise time even under low laser power in our simulations. It is important to note that the simulations assumed direct laser irradiation on the tumor cells, which may not fully account for other factors in practical scenarios, such as light absorption and scattering by the skin. Consequently, the rise time observed in real-world applications could be longer than predicted in our study.

While these results provide promising initial insights into the potential material of 2D MoGe$_2$P$_4$ for photothermal therapy (PTT), further investigation, including clinical trials, is essential to validate its efficacy and safety under physiological conditions. Despite these limitations, our findings provides significant insights of 2D MoGe$_2$P$_4$ as a potential candidate for PTT applications, laying the groundwork for future experimental studies.

# Conclusion

This study demonstrates the exceptional optical and thermodynamic properties of the novel two-dimensional material MoGe$_2$P$_4$. Using computational simulations, we validated MoGe$_2$P$_4$'s superior optical absorption in the NIR-I biological window with a peak near 808 nm, high thermal conductivity, and efficient photothermal conversion, with a temperature rise of 24.8°C with respect to human body temperature after only two minutes of low-power 808 nm laser irradiation. The 2D MoGe$_2$P$_4$'s ability to achieve significant temperature elevation with minimal laser power underscores its promise as a material for photothermal therapeutic applications. Furthermore, the stability and biocompatibility of 2D MoGe$_2$P$_4$ highlight its potential for safe biomedical applications. These simulation study on 2D material serve as groundwork for future experimental validations and clinical investigations to translate these insights into practical therapeutic applications.

# Author's Contribution

Sudipta Saha contributed to the conceptualization, methodology, visualization, software, investigation, drafting of the original manuscript, and editing of the manuscript. Arpan Sur contributed to conceptualization, methodology, visualization, software, investigation, drafting of the original manuscript, and editing the manuscript. Labonno Saha contributed to the analysis and interpretation of the results, the drafting of the original manuscript, and the editing of the manuscript. Md. Kawsar Alam was involved in conceptualization, methodology, visualization, resource management, original draft writing, reviewing and editing the manuscript, and provided supervision throughout the research process.

# Data Availability

The data supporting this article have been included as part of the ESI.

# Conflict of Interest

There are no conflicts to declare.